\documentclass[aps,prb,twocolumn,groupedaddress,showpacs]{revtex4}
\usepackage[utf8]{inputenc}
\usepackage{graphicx}
\usepackage{color}
\usepackage{amsmath}
\usepackage{amssymb}
\usepackage{xcolor}
\usepackage{soul}
\usepackage{float}

\usepackage{hyperref}
\hypersetup{
    colorlinks=true, 
    linktoc=all,     
    citecolor=blue,
    filecolor=black,
    linkcolor=blue,
    urlcolor=blue
}

\begin{document}

\title{Strain, Anharmonicity and Finite-Size Effects on the Vibrational Properties of Linear Carbon Chains}

\author{Grazi\^{a}ni Candiotto$^{1}$\footnote{gcandiotto@iq.ufrj.br}}
\author{Fernanda R. Silva$^{1}$}
\author{Deyse G. Costa$^{2}$ }
\author{Rodrigo B. Capaz$^{1,3}$\footnote{capaz@if.ufrj.br}}
\affiliation{$^{1}$ Instituto de F\'{i}sica, Universidade Federal do Rio de Janeiro, Rio de Janeiro, RJ, 21941-972, Brazil.}
\affiliation{$^{2}$ Departamento de Qu\'{i}mica, Universidade Federal de Vi\c{c}osa, Vi\c{c}osa, MG, 36570-900, Caixa Postal 216, Brazil}
\affiliation{$^{3}$ Brazilian Nanotechnology National Laboratory (LNNano), Brazilian Center for Research in Energy and Materials (CNPEM), Campinas, SP, 13083-100, Brazil.}

\date{\today}

\begin{abstract}
\noindent Linear carbon chains (LCCs) are the ultimate 1D molecular system and they show unique mechanical, optical and electronic properties that can be tuned by altering the number of carbon atoms, strain, encapsulation, and other external parameters. In this work, we probe the effects of quantum anharmonicity, strain and finite size on the structural and vibrational properties of these chains, using high$-$level density functional theory (DFT) calculations. We find strong anharmonicity effects for infinite chains, leading to ground$-$state nuclear wavefunctions that are barely localized at each of the dimerized geometries, \textit{i.e.} strong tunneling occurs between the two minima of the potential energy surface. This effect is enhanced for compressive strains. In addition, vibrational C$-$band frequencies deviate substantially from experimental measurements in long chains encapsulated in carbon nanotubes. On the other hand, calculations for finite chains suggest that quantum anharmonicity effects are strongly suppressed in finite system, even in the extrapolation to the infinite case. For finite systems, vibrational C$-$band frequencies agree well with experimental values at zero pressure. However, these frequencies increase under compressive strain, in contradiction with recent results. This contradiction is not resolved by adding explicitly the encapsulating carbon nanotubes to our calculations. Our results indicate that LCCs embody an intriguing 1D system in which the behavior of very large finite systems do not reproduce or converge to the behavior of truly infinite ones.  
\end{abstract}


\maketitle

\section{Introduction}\label{sec:introduction}
Carbon is one the most abundant element in nature and it can rearrange its valence electrons to form bonds in various types of hybridizations, giving rise to a wide variety of molecular and periodic structures.\cite{casari2016} The different $sp^{n}$ ($n =$ 1, 2, and 3) hybrid orbitals result in molecular structures with linear (1D), planar (2D) or tetrahedral (3D) spatial organization, respectively. 

Carbon materials have been extensively explored in recent decades precisely due the diversity of mechanical, optical and electronic properties that can be obtained through their diverse dimensionality, hybridization, chirality and topology. Early works in carbon nanoscience were inspired from interstellar space research and culminated in the discovery of fullerenes (C$_{60}$),\cite{kroto1985} followed by carbon nanotubes (CNTs),\cite{iijima1991} graphene,\cite{geim2010} as well as the one$-$dimensional (1D) linear carbon chains (LCCs).\cite{Zhao2003} In this sense, LCCs are truly unique, as they consist of the purely $sp-$hybridized allotrope of carbon\cite{chalifoux2010} and they can be considered as the ultimate realization of a 1D system. 

An infinite LCC, also called carbyne, may exhibit semiconducting or metallic behavior, associated to polyyne or polycumulene structures, respectively. In polyyne, bonds alternate between single and triple ($\cdots -$ C $\equiv$ C $-$ C $\equiv$ C $- \cdots$),\cite{kutrovskaya2021} while in polycumulene the chains are composed of successive double bonds ($\cdots =$ C $=$ C $=$ C $=$ C $= \cdots$).\cite{costa2021} Theoretical calculations suggests that the polyynic structure is energetically more favorable than the cumulenic structure.\cite{artyukhov2014,romanin2021,yang2007,yang2008,martinati2022} Experimentally, this information is not so easily accessible, as LCCs are extremely unstable and reactive at ambient conditions.\cite{chalifoux2010} Until recently, the longest polyyne synthesized and stabilized contained 44 carbon atoms.\cite{chalifoux2010} These samples are certainly far from the infinite carbyne limit, and properties are expected to be very much influenced by finite$-$size effects. In addition, geometry determination is often indirect, based on optical and vibrational spectroscopies. 

As single and multiwall CNTs (MWCNT) have the interesting capability of encapsulating different species,\cite{li2021} this feature allowed to use these structures as nanoreactors to synthesize long and stable LCCs at ambient conditions with $>$ 6000 atoms as obtained by Shi \textit{et al.}\cite{shi2016,shi2017}. For these samples, high$-$resolution transmission electron microscopy (HRTEM) revealed directly single$-$triple bond length alternation (BLA $= R_{s} - R_{t}$, the difference between the length of single and triple bonds) compatible with the polyyne structure. In addition, one would expect that, for such long molecules, end$-$group and finite size effects would be negligible. Since then, LCCs encapsulated in MWCNTs (LCC@MWCNT) has been the best alternative to study these materials.\cite{malard2007,aguiar2011,aguiar2012}

Raman spectroscopy is also a powerful technique to study LCCs, as the Raman signal of carbon chains is very intense and provides information on structural changes, chain length and stability.\cite{martinati2022,fantini2006,malard2007,moura2009,andrade2015-2,zirzlmeier2020,marabotti2022} The bond$-$stretching vibrational mode of polyynes (frequently called C$-$band) is easily identified inside MWCNT by its Raman fingerprint frequency $\omega_{LCC}$, which lies in the 1820 $-$ 1870 cm$^{-1}$ spectral region at ambient conditions.\cite{andrade2015-1,neves2018,moura2023} It is also well known that $\omega_{LCC}$ is proportional to $N^{-1}$, where $N$ is the number of carbon atoms. Recently, Sharma \textit{et. al.}\cite{sharma2020} reported that polyynes@MWCNT when subjected to pressures up to 4.60 GPa shows a reversible $\omega_{LCC}$ softening associated with linear redshift as large as 22 cm$^{-1}$, in agreement with previous measurements.\cite{andrade2015-1,neves2018} This is quite unusual since pressure typically induces frequency hardening in $sp^{2}$ carbon materials.\cite{sandler2003,alvarez2010,proctor2009} Coalescence and charge transfer effects have been suggested as the driving mechanism for this behavior.\cite{andrade2015-1,neves2018} An empirical model based on single and triple bond anharmonicities has also been proposed.\cite{sharma2020} As a matter of fact, quantum anharmonicity effects have been predicted to play a key role in the structural, electronic and optical properties of carbyne,\cite{romanin2021,artyukhov2014} particularly in the understanding of the metal$-$to$-$insulator (polycumulene$-$to$-$polyyne) transition that can be in principle driven by pressure and temperature effects.  

In this work, we further explore quantum anharmonicity in combination with strain and finite$-$size effects in the structural and vibrational properties of infinite and finite LCCs. Our results indicate that quantum anharmonicity effects are much more relevant in infinite systems, as compared to finite ones. Surprisingly, the vibrational properties of finite LCCs in the $N \to \infty$ limit do not seem to converge to the truly infinite behavior, as end effects appear not the disappear in this limit. We find that C$-$band frequencies increase with compressive strain both for infinite and finite LCCs, in apparent disagreement with experiments.\cite{sharma2020}

\section{COMPUTATIONAL DETAILS}
It is well$-$known that a high$-$level treatment of exchange$-$correlation effects is required to properly reproduce the correct structural and vibrational properties of LCCs $-$ specially infinite ones $-$ using DFT.\cite{ramberger2021,jacquemin2006,romanin2021,candiotto2020} In this work, we adopt the hybrid exchange$-$correlation functional mCAM$-$B3LYP,\cite{day2006} that combines the hybrid qualities of Becke three$-$parameter Lee$-$Yang$-$Parr (B3LYP)\cite{becke1988,LYP1988} and Coulomb$-$attenuating method (CAM)\cite{yanai2004} using the long$-$range correction parameters $\beta=0.19$ (keeping $\alpha=0.19$ and $\mu=0.33$). The mCAM$-$B3LYP functional provides a good description of structural properties of LCCs,\cite{romanin2021} particularly of the BLA, which is a key parameter that determines many other structural, optical, vibrational and electronic properties of dimerized systems.\cite{yang2007,yang2008}

\begin{figure}[b]
    \centering
    \includegraphics[width=\linewidth]{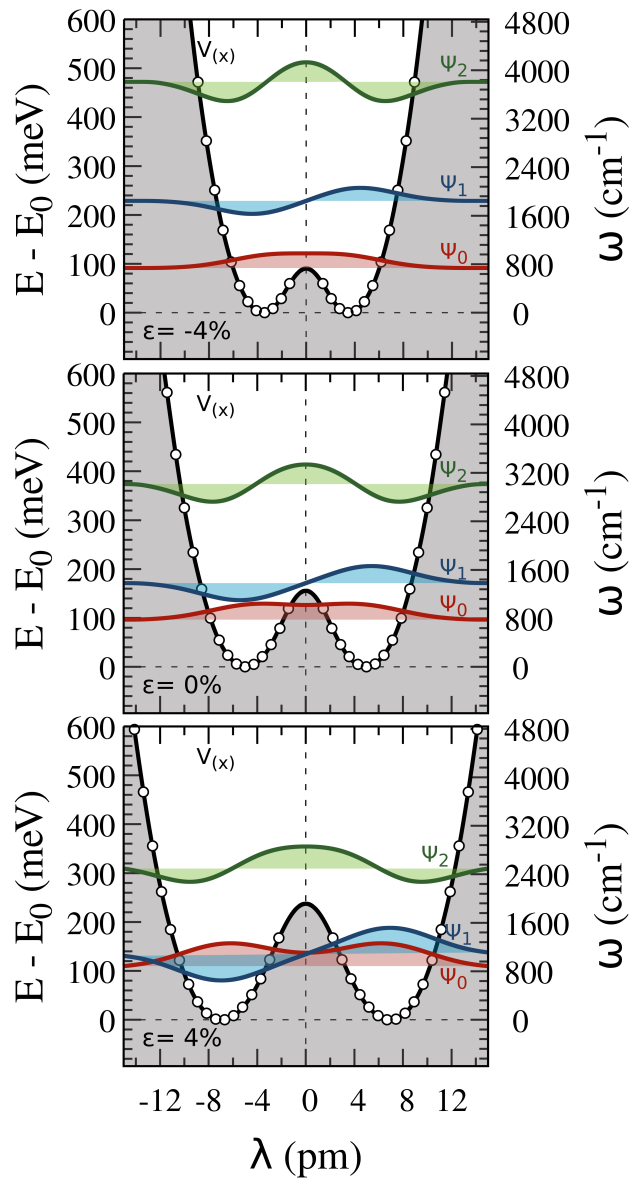}
    \caption{Total (potential) energy (dots) for carbyne in different strain conditions. Solid black lines represents the fits using a 16$^{th}-$order polynomial. Colored lines are the three lowest energy levels and their respective wavefunctions, with baselines positioned at their respective eigenvalues.}
    \label{fig:panel-wf-carby}
\end{figure}

To explore finite size effects, we perform calculations for both carbyne and polyynes. To model carbyne we use the \textit{Quantum Espresso}\cite{giannozzi2009} package along with the library of exchange$-$correlation and kinetic energy functionals \textit{LIBXC}.\cite{marques2012,lehtola2018} These calculations are based on plane$-$wave basis sets, norm$-$conserving pseudopotentials,\cite{hamann2013} k$-$points sampling of 1$\times$1$\times$100 and a plane$-$wave energy cutoff of 50 Ry. A tetragonal unit cell was constructed, assuming the infinite polymer chain along the $c$ axis and an interchain distance of 10 \AA. In the description of polyynes and polyynes encapsulated by a single$-$wall carbon nanotube (C$_{10}$H$_{2}$@SWCNT) the calculations were based on a linear combination of atomic orbitals (LCAO) where the functional mCAM$-$B3LYP was also used along with a split$-$valence double$-$zeta polarized (6$-$31G (d, p))\cite{pople1971} basis set based in Gaussian type orbitals (GTOs). In the calculations for C$_{10}$H$_{2}$@SWCNT, we also included Grimmes’s dispersion correction (GD3)\cite{grimme2010,grimme2011} to improve the description of van der Waals interactions. These calculations were performed using the \textit{ORCA}\cite{ORCA2012,ORCA2022} package also together with the \textit{LIBXC} library. We use LCAO methods for encapsulated systems because LCC and SWCNT unit cells are incommensurate. Therefore, one would have to use a prohibitively large supercell along the chain and nanotube axis in order do construct a stress$-$free system in which both subsystems are relaxed to their equilibrium geometries. For LCAO codes, this is not a problem, because they do not rely on periodic boundary conditions. Comparative validation of plane$-$wave and LCAO methods was achieved by performing calculations of the BLA in the center of isolated C$_{10}$H$_2$ chains. The calculated BLA using both methods differ by only 2\%.

Raman scattering involves an inelastic process in which a photon loses or gains energy to a vibrational excitation, regardless of whether this vibration is harmonic or not. Therefore, the proper way to compare Raman spectra with theory for an anharmonic potential is through the calculation of (anharmonic) energy levels and by comparing the interlevel energy difference with the position of Raman peaks. In the particular case of harmonic approximation, vibrational frequencies were directly calculated by both \textit{Quantum Espresso} and \textit{ORCA} codes. In the anharmonic case, vibrational levels and wavefuntions were obtained by directly solving the Schr\"{o}dinger equation\cite{schrodinger1926} for the nuclei in the Born$-$Oppenheimer energy surface of a single normal$-$mode (C$-$band) coordinate. In this approximation, coupling between the C$-$band and other modes is neglected.\cite{artyukhov2014}

\section{RESULTS AND DISCUSSION}
\subsection{Infinite LCC (carbyne)}
Fig. \ref{fig:panel-wf-carby} shows the calculated total energy (per unit cell) curves as a function or the C$-$band normal$-$mode coordinates $\lambda$ for various strain values. The coordinate $\lambda$ relates to the BLA as $\lambda=BLA/\sqrt{8}$.\cite{wilson2019} As expected, a symmetric and anharmonic potential is obtained from the calculations. The two minima correspond to the two equivalent dimerized structures, which correspond to an exchange between triple and single bonds,\cite{pereira2020,artyukhov2014} while the top of potential barrier corresponds to the equally$-$spaced double bond (cumulene) structures. For an infinite system, both ground states are equivalent and present a closed$-$shell electronic structure with an energy gap between occupied and empty states (Peierls gap). The Peierls barrier between them (energy difference between cumulene and polyyne structures) is 155 meV per unit cell (78 meV/atom), to be compared to 34 meV/atom obtained using the PBE0 functional\cite{romanin2021} and $\approx$ 40 meV using the HSE06 functional.\cite{artyukhov2014} Despite the larger mCAM$-$B3LYP barrier, tunneling between the two minima is still strong: At zero strain, zero$-$point energy is slightly below the barrier and the ground state wavefunction is just barely double$-$peaked, indicating that the classical picture of carbyne ground$-$state as a dimerized structure is a severe (if not incorrect) approximation. As pointed out by Artyukhov,\cite{artyukhov2014} the Peierls barrier and the probability amplitudes for dimerized structures increase for positive strains (extension) and decrease for negative ones (compression).  

 \begin{figure}[b]
     \centering
     \includegraphics[width=\linewidth]{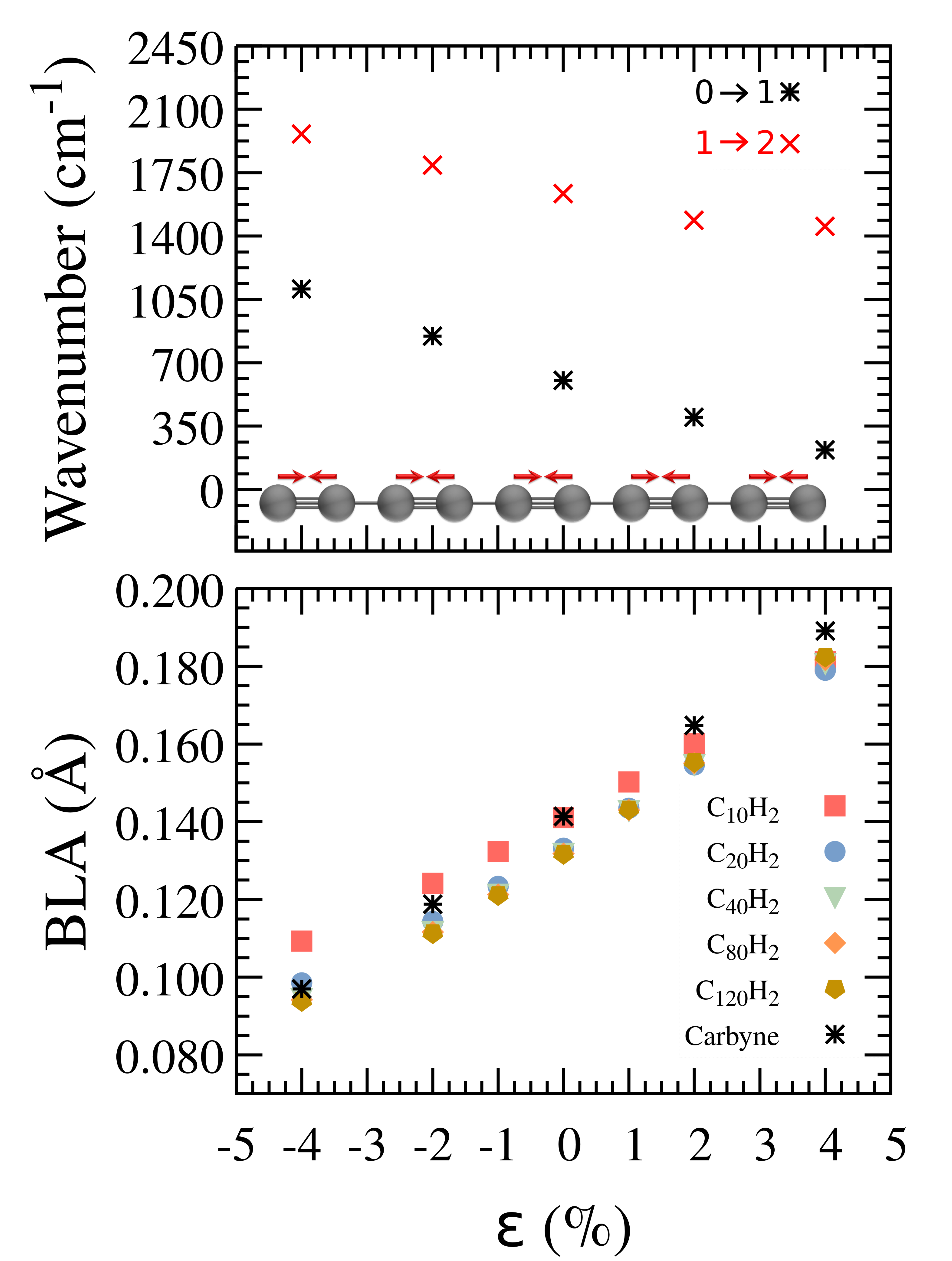}
     \caption{C-band $0\to1$ and $1\to2$ frequencies in different strain conditions (top). The ball$-$and$-$stick model in inset indicates the normal mode displacements of the C$-$band. Bond length alternation (BLA) for carbyne and polyynes under strain (bottom).}
     \label{fig:BLA}
 \end{figure}

We now turn our attention to the vibrational levels. We solve the single$-$mode Schr\"{o}dinger equation with the full anharmonic potential and the resulting eigenvalues and wavefunctions are shown in Fig. \ref{fig:panel-wf-carby}. We focus attention on vibrational transitions between the ground state and the first excited state ($0\to1$ transitions) and from the first to the second excited states ($1\to 2$ transitions). First of all, notice that the $0\to1$ and $1\to2$ frequencies are very different from each other (see top panel of Fig. \ref{fig:BLA}), as expected for a strongly anharmonic potential: At zero strain, $\omega_{0\to1} =$ 603 cm$^{-1}$ and $\omega_{1\to2} =$ 1634 cm$^{-1}$. Therefore, the $0\to1$ frequency (which is expected to dominate the Raman spectra at reasonably low temperatures) is substantially different than the experimental values. Experimentally, we find 1792 cm$^{-1}$, by a quadratic extrapolation to $N\to\infty$ of the experimental data from Kastner \textit{et al.}\cite{kastner1995}) and 1791 cm$^{-1}$ for the longest LCC (797 nm) encapsulated in a carbon nanotube.\cite{shi2016} The top panel of Fig. \ref{fig:BLA} shows the strain dependence of both  $\omega_{0\to1}$ and $\omega_{1\to2}$. One can see that the usual behavior of frequencies increasing upon compressive strain is not changed by the addition of quantum anharmonic effects.

In summary, our calculations for the infinite LCC challenge the picture of a dimerized ground state and they are apparently irreconcilable with the experimental values of vibrational frequencies. In the following, we'll shed light on this apparent discrepancy by performing calculations for finite LCCs.

\subsection{Finite LCCs (polyynes)}

\begin{figure}[b]
    \centering
    \includegraphics[width=\linewidth]{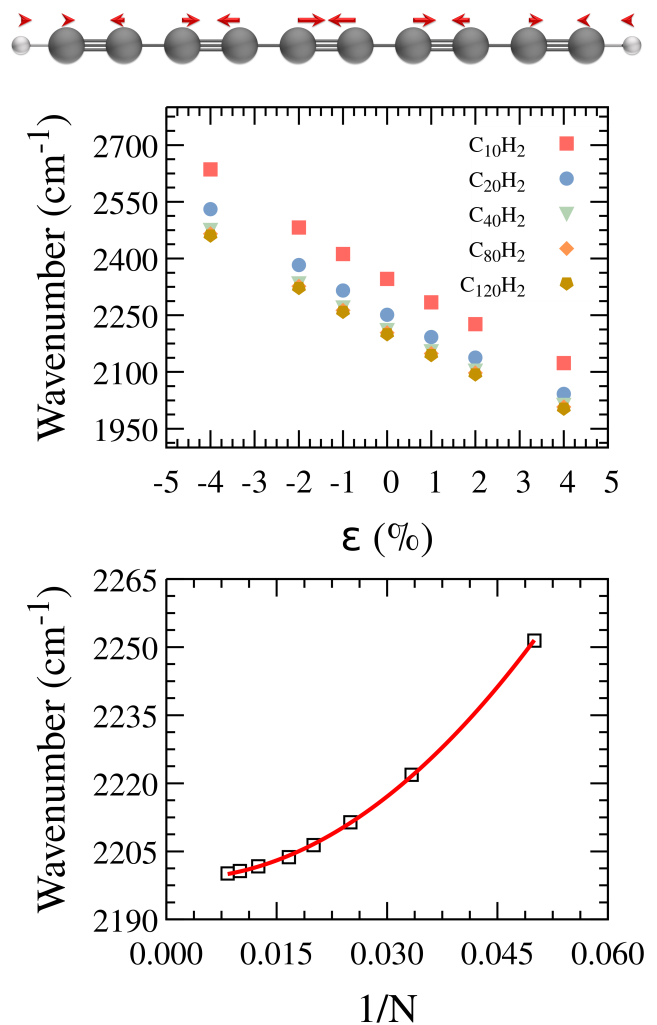}
    \caption{The ball$-$and$-$stick model indicates the normal mode displacements of the C$-$band for C$_{10}$H$_{2}$ structure (top). C$-$band frequency for polyynes of different sizes as function of strain (middle). Behaviour of C$-$band frequency as function of inverse of number of carbon atoms (bottom).}
    \label{fig:panel-freq}
\end{figure}

We now address finite size effects by performing calculations on polyynes of different sizes (from $N=10$ to $N=120$, where $N$ is the number of carbon atoms in the chain, terminated by hydrogen atoms (C$-$H groups) on both ends. Fig. \ref{fig:BLA} (bottom) shows the BLA at the center of the chains as function of strain for carbyne and polyynes with different sizes. In the case of finite polyynes, strain is applied by constraining the positions of H atoms at chain ends. Notice that BLA of carbyne and polyynes have the same qualitative behavior under strain, decreasing for compression and increasing for tension. For zero strain, we find BLAs of 0.141 and 0.132 {\AA} for carbyne and polyyne ($N=120$) respectively, in agreement with the literature for similar types of exchange-correlation potentials.\cite{romanin2021,ramberger2021}

\begin{figure}[b]
    \centering
    \includegraphics[width=\linewidth]{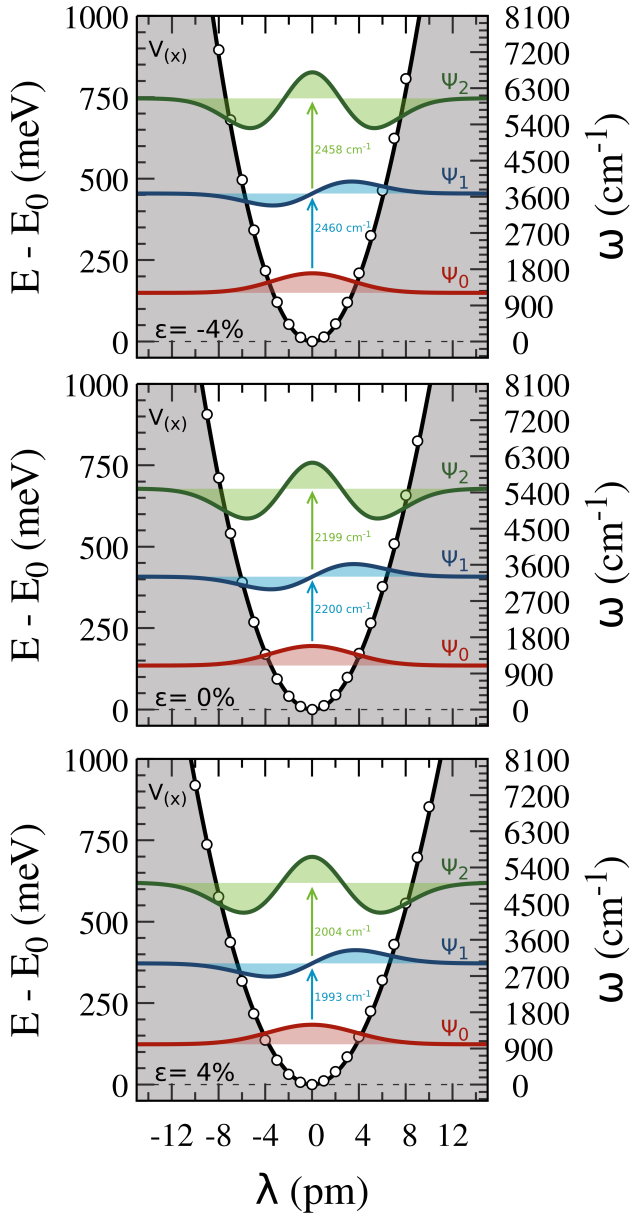}
    \caption{Energy surface potential for polyyne (C$_{120}$H$_{2}$) in different strain conditions. Solid black lines represents the fit of Eq. (\ref{eq:morse}), while solid colored lines are the three lowest energy levels and its respective wavefunctions for each strain conditions.}
    \label{fig:panel-wf-poly}
\end{figure}

To address anharmonic effects in the vibrational properties of finite polyynes, we first identify the corresponding C$-$band mode using the harmonic approximation and then we calculate the total energy as a function of mode displacement (``frozen$-$phonon" method). Notice that C$-$band mode has zero displacement at the H atoms, as they should correspond in fact to a finite wavelength mode of the infinite polyyne (carbyne) C$-$band phonon branch, calculated in the previous subsection at the $\Gamma-$point ($q=0$). An illustration of bond$-$stretching vibrational mode is shown on the top of Fig. \ref{fig:panel-freq} for C$_{10}$H$_{2}$. The longer the polyyne, the longer the wavelength, and the mode frequency should in principle gradually approach the $\Gamma-$point mode of carbyne as $N\to\infty$. Fig. \ref{fig:panel-wf-poly} shows the total energy as a function of mode displacement for the longest polyyne (C$_{120}$H$_{2}$) studied, under different strain conditions. One clear notices that anharmonic effects are much smaller than in the case of carbyne. As a matter of fact, no double$-$minimum structure is observed and this is clearly a consequence of H$-$termination: Hydrogens impose that the C$-$H bonds at the ends are single bonds and effectively destroy the degeneracy between the two alternating triple$-$single bonds ground states of infinite carbyne. 

As a consequence, the single$-$mode potential energy for finite LCCs is only slighly anharmonic and it can be well described by the Morse potential\cite{morse1929} (solid black lines in Fig. \ref{fig:panel-wf-poly}):

\begin{equation}\label{eq:morse}
V(x) = D_{e}\left( 1 - \exp\left(-ax\right)\right)^{2},
\end{equation}

\noindent where $D_{e}$ is the dissociation energy, $a=\sqrt{k_e/2D_{e}}$ and $k_{e}=\left(d^{2}V/dx^{2}\right)_{e}$ is the force constant at the bottom of the potential well. The three lowest energy level and their wavefunctions in different strain conditions can be obtained analytically and they are shown in Fig. \ref{fig:panel-wf-poly}. Since the potential is only slightly anharmonic,  $\omega_{0\to1}$ and $\omega_{1\to2}$ are very similar to each other. 

In Fig. \ref{fig:panel-freq} (middle) we plot $\omega_{0\to1}$ as a function of strain. For all chain lengths, vibrational frequencies increase upon compression. To understand how the chain length affect the C$-$band frequency, Fig. \ref{fig:panel-freq} (bottom) shows the dependence of $\omega_{0\to1}$ as function of the inverse of number of carbon atoms ($1/N$) in the chain, for $\varepsilon$ = 0\%. We observe a quadratic dependence of $\omega_{0\to1}$ as function of $1/N$: $\omega_{0\to1}=\omega_{\infty}+A/N+B/N^2$, where $\omega_{\infty}=2192$ cm$^{-1}$, $A=674$ cm$^{-1}$ and $B=8788$ cm$^{-1}$. In agreement with the literature,\cite{romanin2021} the mCAM$-$B3LYP functional seems to overestimate the vibrational frequencies not only for the infinite chain, but also for finite LCCs.  

It is striking that the C$-$band vibrational frequency of finite LCCs extrapolated to $N\to\infty$ (2192 cm$^{-1}$) is completely different from the calculated value for the truly infinite chain (603 cm$^{-1}$ for $\omega_{0\to1}$). It is generally believed that the behavior of finite systems in the limit of sizes approaching infinity should reproduce the behavior of truly infinite systems with periodic boundary conditions, as surface (or end) effects should gradually become negligible. It is evident that this is not the case here, where end effects imposed by C$-$H terminations propagate through the whole chain, destroying the double$-$degeneracy of dimerized states and washing out the effects of quantum anharmonicity.

\begin{figure}[H]
    \centering
    \includegraphics[width=\linewidth]{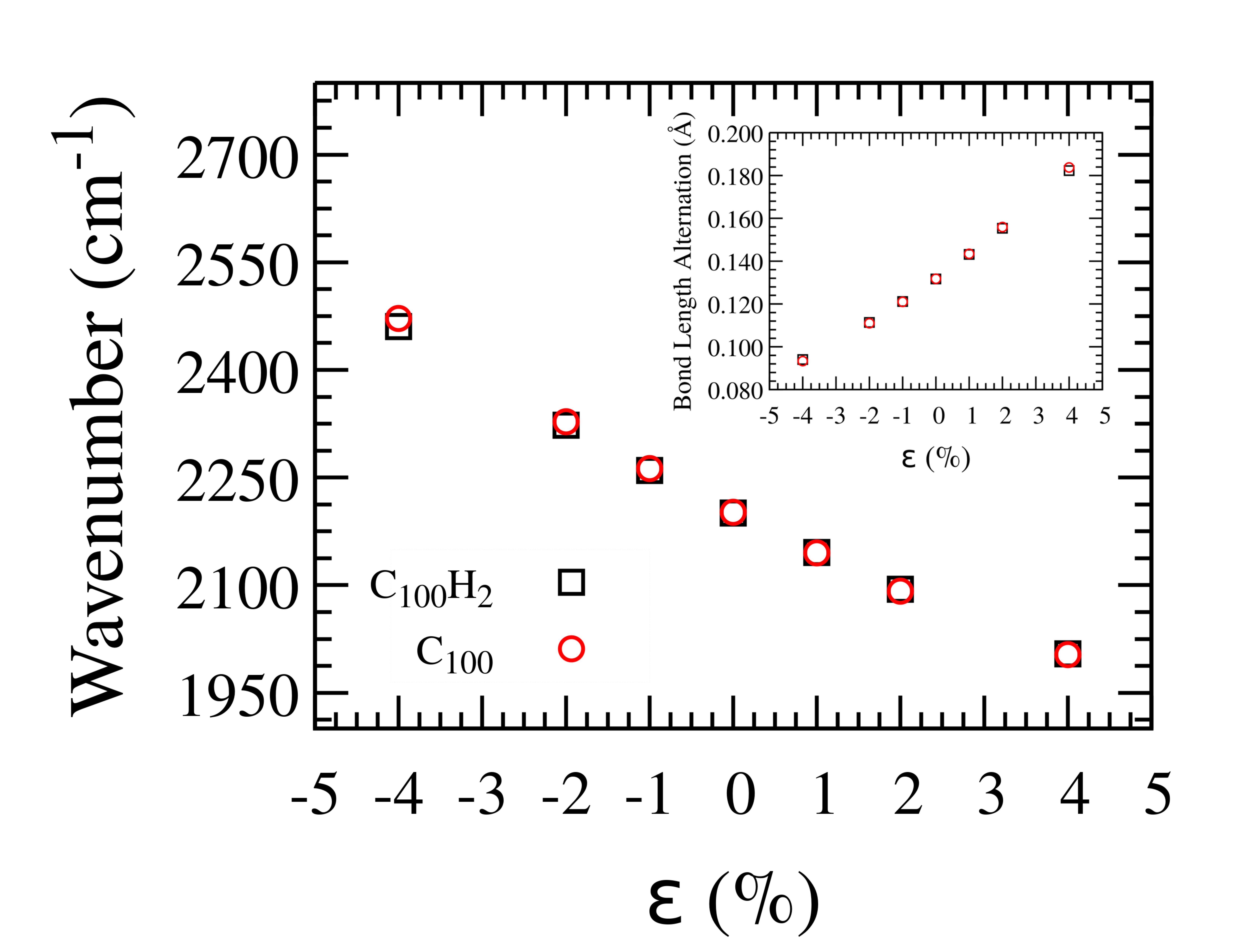}
    \caption{Comparison between vibrational properties of C$_{100}$H$_{2}$ (saturated) and C$_{100}$ (unsaturated) LCCs under extension and compression. The inset shows the behavior of the BLA for C$_{100}$H$_{2}$ and C$_{100}$ LCCs.}
    \label{fig:C100-comp}
\end{figure}

Finally, we address the possible impact of the particular type of chain termination by H atoms adopted in this work on the results of BLA and C$-$band frequency of finite LCCs. We do not expect that H-terminated or unsaturated chains behave in different ways, simply because both terminations impose that the ending carbon bond is single, therefore breaking the doubly$-$degenerate ground state structure of the infinite chain. The only difference is the presence of a saturated C$-$H bond in one case and a dangling single bond in the other case. Other than producing localized states near the HOMO$-$LUMO gap, the dangling bonds should have no effect whatsoever on the vibrational properties of sufficiently long chains. To prove that, we perform calculations on an unsaturated LCC containing 100 carbon atoms (C$_{100}$). Figure \ref{fig:C100-comp} shows a comparative analysis, illustrating the results for C$-$band frequency and BLA in the center of the chain (inset) under strain, comparing the results of saturated (C$_{100}$H$_{2}$) and unsaturated (C$_{100}$) finite LCCs. As expected, the results for the BLA and C$-$band vibration frequency of finite LCCs are not influenced by H$-$termination.

\subsection{Polyynes@single-wall CNT}

\begin{figure}[H]
    \centering
    \includegraphics[width=\linewidth]{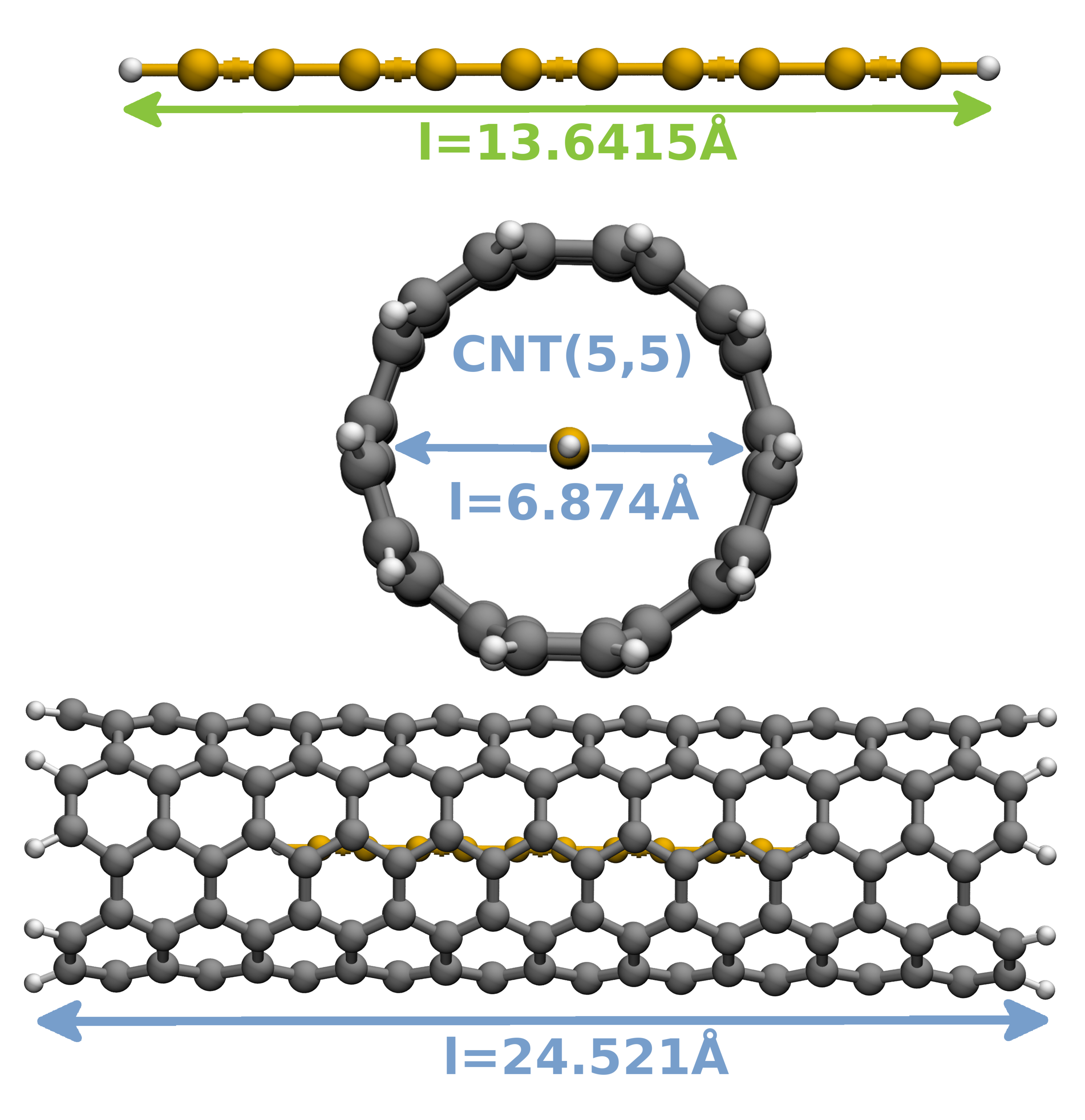}
    \caption{Illustration of C$_{10}$H$_{2}$@SWCNT(5,5) used in this work.}
    \label{fig:cnt-panel}
\end{figure}

In a last effort to trying to understand the anomalous results of Sharma \textit{et al.}\cite{sharma2020} regarding C$-$band softening under pressure, we now explicitly introduce encapsulation effects due to carbon nanotubes. In this way, we can probe the effects of radial component of pressure (rather than a pure axial strain, as described in the two previous subsections) and also investigate possible charge transfer and chemical interactions. We consider a short polyyne C$_{10}$H$_{2}$ encapsulated by a finite segment of SWCNT(5,5), as shown in Fig. \ref{fig:cnt-panel}. First, the molecular structure of C$_{10}$H$_{2}$@SWCNT(5,5) was relaxed to the ground state geometry. Fig. \ref{fig:cnt-panel} shows the geometry of C$_{10}$H$_{2}$@SWCNT(5,5) after the relaxation process. The structure is then submitted to a purely radial pressure and the system is relaxed again. Pressure was applied via external radial forces on SWCNT atoms according to the prescription of Capaz \textit{et al.} \cite{capaz2004}$^{,}$\footnote{Axial strain under pressure was neglected since SWCNTs are much softer radially than axially under hydrostatic pressure\cite{capaz2004}. Indeed, we performed tests in which both radial and axial strains are properly included and the difference is negligible with respect to the case of purely radial strain.}. This procedure was performed for different pressures and after the relaxation process the polyyne remained always in the center of the SWCNT and no chemical bond between polyyne and the walls of the CNT was formed. 

\begin{figure}[H]
    \centering
    \includegraphics[width=\linewidth]{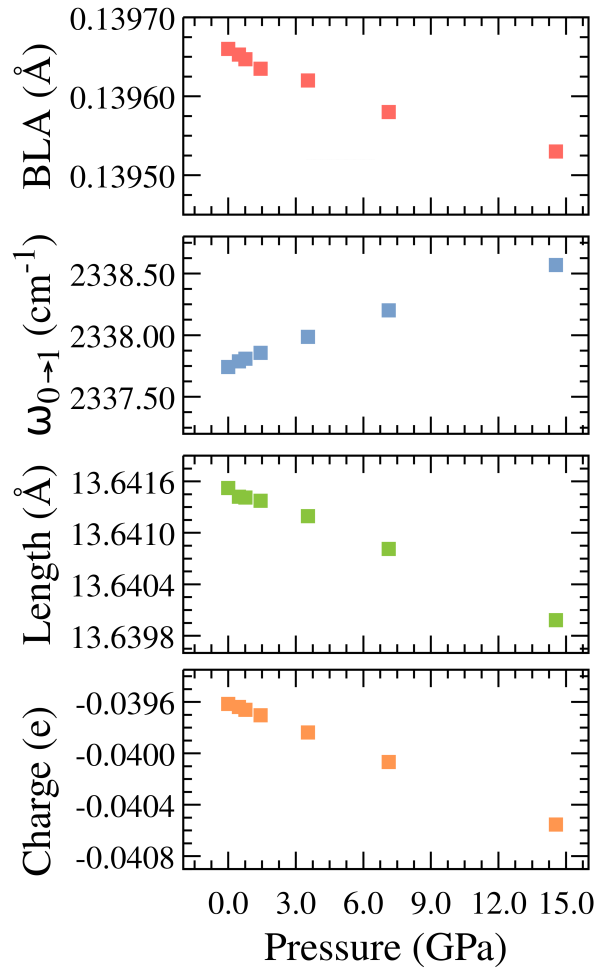}
    \caption{Bond length alternation (top) and C$-$band frequency as function of pressure for C$_{10}$H$_{2}$ encapsuled by SWCNT(5,5) (bottom).}
    \label{fig:cnt-results}
\end{figure}

Fig. \ref{fig:cnt-results} shows the behavior of BLA, $\omega_{0\to1}$, polyyne length (as measured from the distance between H atoms) and transferred charge as function of pressure. Notice that the polyyne length shortens under pressure, even though the pressure is purely radial, as if polyyne had a negative Poisson ratio at the molecular level. As a consequence, the system behaves qualitatively in a similar manner as the isolated polyyne under axial compressive strain, described in the previous subsection: The BLA decreases and the C$-$band vibrational frequency increases under pressure. Charge transfer effects appear to be very small, in consistency with the fact that the Fermi level of the metallic (5,5) SWCNT is positioned within the HOMO$-$LUMO gap of C$_{10}$H$_{2}$. 

\section{Conclusions}\label{conc}
In conclusion, we used DFT with high$-$level exchange$-$correlation functionals to described the combined effects of finite$-$size, quantum anharmonicity and strain on the structural and vibrational properties LCCs. Anharmonicity is strong for infinite systems but not for finite ones, as the effects of C$-$H terminations propagate throughout the chain and suppress the double$-$minima structure of the potential energy surface that occur for infinite systems. As a consequence, the physics of truly infinite systems is not reproduced by an $N\rightarrow \infty$ extrapolation of finite ones, as end effects do not disappear in this limit. Experimental vibrational frequencies are only reproduced by the calculations of finite systems, thus indicating that a truly infinite LCC may be elusive to experimental realization. Compressive strain increase C$-$band vibrational frequencies for both infinite and finite systems (with and without SWCNT encapsulation), in apparent disagreement with experiments.    

\section{Acknowledgements}
The authors acknowledge financial support from Brazilian agencies CNPq, FAPERJ, FAPESP, FINEP, INCT $-$ Carbon Nanomaterials, INCT $-$ Materials Informatics and INCT $-$ INEO for financial support. G.C. gratefully acknowledge FAPERJ, grant number E-26/200.627/2022 and E-26/210.391/2022 (project Jovem Pesquisador Fluminense process number 271814) for financial support. The authors also acknowledge the computational support of N\'{u}cleo Avan\c{c}ado de Computa\c{c}\~ao de Alto Desempenho (NACAD/COPPE/UFRJ) and Sistema Nacional de Processamento de Alto Desempenho (SINAPAD). 

\bibliographystyle{apsrev4-2}

\bibliography{lcc-bib}

\end{document}